\documentclass[conference,a4paper,hidelinks=false]{IEEEtran}
\usepackage{amsmath}
\usepackage{amssymb}
\usepackage{booktabs}
\usepackage{cite}
\usepackage{graphicx}
\usepackage{array}
\usepackage{multirow}
\usepackage[caption=false,font=footnotesize]{subfig}
\usepackage{xcolor}
\usepackage{orcidlink}
\usepackage{hyperref}
\hypersetup{
    colorlinks=true,
    linkcolor=blue,
    citecolor=blue,
    urlcolor=blue
}

\newcommand{\ours}{MambaRefine-CD}
\newcommand{\drbi}{D-RBI}

\newcommand{\best}[1]{\textcolor{red}{\textbf{#1}}}
\newcommand{\second}[1]{\textcolor{blue}{\textbf{#1}}}
\newcommand{\third}[1]{\textcolor{teal}{\textbf{#1}}}

\title{MambaRefine-CD: MambaVision with Region-Boundary Temporal Refinement}

\author{
\IEEEauthorblockN{
\begin{minipage}{0.78\textwidth}
\centering
\begin{tabular*}{\textwidth}{@{\extracolsep{\fill}}ccc}
\begin{tabular}{c}
Dineth Perera\orcidlink{0009-0009-1295-2547}\\
\textit{University of Peradeniya}\\
Peradeniya, Sri Lanka\\
e21291@eng.pdn.ac.lk
\end{tabular}
&
\begin{tabular}{c}
Thaariq Firdous\orcidlink{0009-0005-6150-2228}\\
\textit{University of Peradeniya}\\
Peradeniya, Sri Lanka\\
e21139@eng.pdn.ac.lk
\end{tabular}
&
\begin{tabular}{c}
Oshadha Samarakoon\orcidlink{0009-0000-9337-0198}\\
\textit{University of Peradeniya}\\
Peradeniya, Sri Lanka\\
e21345@eng.pdn.ac.lk
\end{tabular}
\end{tabular*}
\end{minipage}
}

\vspace{0.4cm}

\IEEEauthorblockN{
\begin{minipage}{0.78\textwidth}
\centering
\begin{tabular*}{\textwidth}{@{\extracolsep{\fill}}ccc}
\begin{tabular}{c}
Roshan Godaliyadda\orcidlink{0000-0002-3495-481X}\\
\textit{University of Peradeniya}\\
Peradeniya, Sri Lanka\\
roshang@eng.pdn.ac.lk
\end{tabular}
&
\begin{tabular}{c}
Parakrama Ekanayake\orcidlink{0000-0002-5639-8105}\\
\textit{University of Peradeniya}\\
Peradeniya, Sri Lanka\\
mpbe@eng.pdn.ac.lk
\end{tabular}
&
\begin{tabular}{c}
Vijitha Herath\orcidlink{0000-0002-2094-0716}\\
\textit{University of Peradeniya}\\
Peradeniya, Sri Lanka\\
vijitha@eng.pdn.ac.lk
\end{tabular}
\end{tabular*}
\end{minipage}
}
}
\begin{document}
\maketitle

\begin{abstract}
Binary change detection in remote sensing requires both complete changed-region localization and accurate boundary delineation. We present MambaRefine-CD, a region-boundary temporal refinement framework built on a shared MambaVision encoder. The proposed D-RBI module constructs temporal evidence from paired features, absolute differences, and signed differences, then separates it into region and Sobel-conditioned boundary streams. Region features are enhanced with CRAM-lite and decoded by an adaptive receptive-field FPN, while the finest boundary stream guides a bounded residual refinement of the coarse prediction. Experiments on DSIFN-CD and WHU-CD show strong changed-class F1 and IoU under verified evaluation settings, and ablations support the contribution of signed temporal evidence and the full region-boundary refinement pipeline.Code available \href{https://github.com/Dineth14/MambaRefine-CD}{here}.
\end{abstract}


\section{Introduction}
\label{sec:introduction}

Remote-sensing change detection (CD) estimates a binary change mask from two co-registered images acquired over the same area at different times. The task underpins urban expansion monitoring, disaster assessment, land-cover analysis, and infrastructure management. In practice, binary CD is difficult because the model must compare two observations while being invariant to nuisance variation from illumination shifts, seasonal changes, shadows, sensor differences, and small registration errors. These effects are most visible near object boundaries, where a model may detect the changed region but still produce shifted, fragmented, or over-smoothed masks.

Deep CD methods have evolved along three main directions. Convolutional Siamese networks compare bi-temporal images through early fusion or feature differencing, as in FC-EF, FC-Siam-conc, and FC-Siam-diff~\cite{daudt2018fully}. Later CNN encoder-decoder models improve localization through skip connections, dense feature reuse, and multi-scale attention~\cite{ronneberger2015unet,fang2022snunet,chen2021dasnet}. These models are effective because convolutions provide useful locality and translation bias. Their context modeling is still bounded by the receptive field, which can limit reasoning about spatially distributed changes. Transformer-based methods address long-range interaction more directly. BIT models bi-temporal tokens with a Transformer encoder~\cite{chen2022bit}, and ChangeFormer uses a hierarchical Siamese Transformer for dense CD~\cite{bandara2022changeformer}. Wijenayake et al.~\cite{precision_spatiotemporal_fusion_iciis2025} further demonstrate that precise spatio-temporal feature fusion yields gains over standard differencing strategies. Despite this progress, attention-based models carry high memory and compute cost at dense resolution, and stronger global representations do not automatically resolve boundary localization errors.

Recent visual state-space models (SSMs) provide a different route. Mamba introduces selective state-space sequence modeling with linear complexity~\cite{gu2023mamba}, and VMamba adapts it to vision through two-dimensional selective scanning~\cite{liu2024vmamba}. MambaVision develops a hybrid Mamba-Transformer backbone for visual recognition and dense prediction tasks~\cite{hatamizadeh2025mambavision}. Mamba-based CD models, including ChangeMamba~\cite{chen2024changemamba}, HAM-CD~\cite{li2026hamcd}, and Mamba-CD~\cite{peng2026mambacd}, show that SSM backbones are competitive for binary and semantic change detection. Wijenayake et al.\ in Mamba-FCS~\cite{mambafcs} further demonstrate that combining frequency-domain fusion with change-guided attention and a SeK-inspired loss improves semantic CD. However, applying a Mamba-based encoder alone is not a complete solution. A CD model must still construct temporal evidence from both images and preserve fine boundaries in the output mask. A controlled benchmark of SSM backbones for remote-sensing segmentation ~\cite{wasalathilaka2026controlled} found that boundary delineation remains the dominant failure mode under domain shift, suggesting that encoder scaling alone is insufficient for boundary-sensitive tasks.

We propose \textbf{\ours}, a MambaVision-based architecture for binary remote-sensing CD. The model builds temporal features using raw paired features, absolute difference, and signed difference, giving the decoder access to both magnitude-based and direction-aware evidence. The core novelty is \textbf{\drbi}, which applies two separate learned gating mechanisms to temporal features before decoding. A region gate produces stable changed-area responses, and a Sobel-conditioned boundary gate selects contour-sensitive evidence. These two pathways are combined through an adaptive receptive-field decoder and refined by a bounded residual head that applies a controlled correction to coarse logits near boundaries.

The contributions of this paper are:
\begin{itemize}
    \item We propose MambaRefine-CD, a MambaVision-based binary change detection framework that combines shared bi-temporal encoding with explicit region-boundary temporal refinement.

    \item We introduce D-RBI, a lightweight differential region-boundary interaction module that constructs temporal evidence from paired features, absolute difference, 
    
\begin{figure*}[t]
\centering
\includegraphics[width=\textwidth]{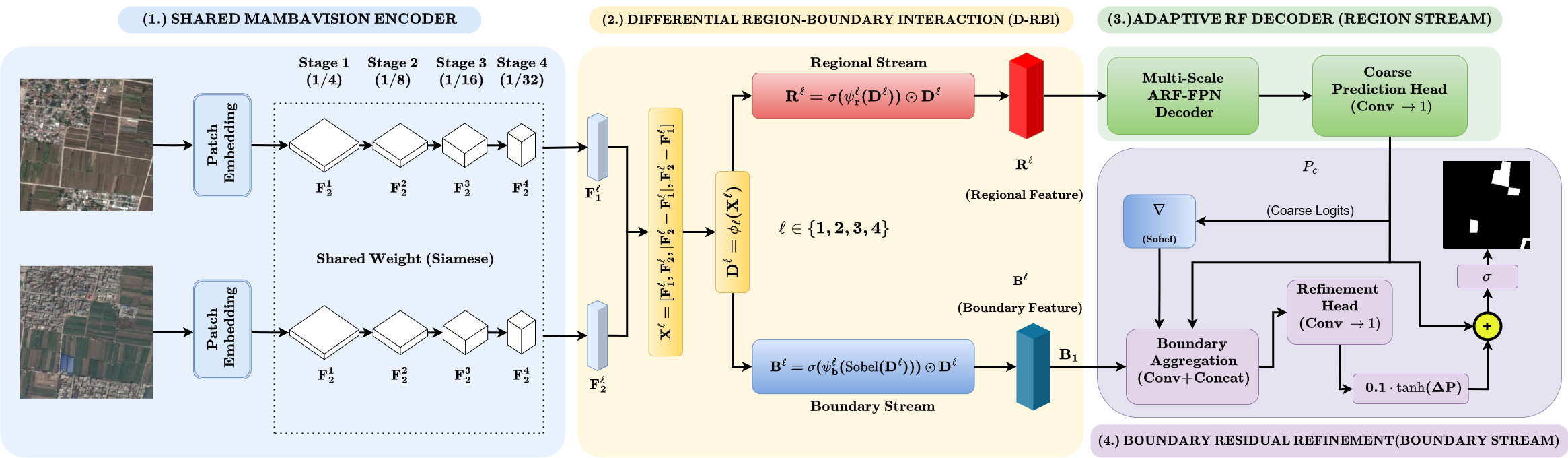}
\caption{Overall architecture of MambaRefine-CD. Two bi-temporal images are processed by a shared MambaVision encoder. Same-scale features are fused by D-RBI modules, region features are decoded by ARF-FPN, and the finest boundary stream guides a bounded residual refinement of the coarse prediction.}
\label{fig:architecture}
\end{figure*}

    and signed difference, and separates the representation into region-aware and boundary-aware streams.

    \item We design a bounded boundary residual refinement head that applies controlled corrections to coarse logits near predicted boundary regions, reducing the risk of overwriting the global changed-region prediction.
    
\end{itemize}

The remainder of the paper is organized as follows. Section~\ref{sec:related} reviews related work. Section~\ref{sec:method} presents the proposed method. Section~\ref{sec:experiments} describes the datasets, metrics, results, ablations, and limitations. Section~\ref{sec:conclusion} concludes the paper.
\section{Related Work}
\label{sec:related}

\subsection{Convolutional and Transformer-Based Change Detection}
Early deep CD models used convolutional Siamese encoders with feature differencing or concatenation for temporal comparison~\cite{daudt2018fully}. Later encoder-decoder models improved localization through skip connections, dense feature reuse, and multi-scale aggregation~\cite{ronneberger2015unet,fang2022snunet,chen2021dasnet}. These CNN methods are efficient and translation-equivariant, but their effective receptive field can limit long-range reasoning over spatially distributed changes.

Transformer-based CD methods address global context more directly. BIT represents bi-temporal features as semantic tokens for temporal reasoning~\cite{chen2022bit}, while ChangeFormer uses a hierarchical Siamese Transformer with an MLP decoder for dense prediction~\cite{bandara2022changeformer}. Ratnayake et al.\ also show that attention and improved loss design can benefit CD architectures~\cite{enhanced_scannet_mercon2025}. However, attention-based models remain expensive at high resolution, and stronger global modeling does not by itself guarantee accurate boundary delineation.

\subsection{State Space Models for Remote Sensing}
Mamba introduces selective SSM sequence modeling with linear complexity~\cite{gu2023mamba}, and VMamba adapts this idea to vision using two-dimensional selective scanning~\cite{liu2024vmamba}. MambaVision further develops a hybrid Mamba-Transformer backbone with strong accuracy-efficiency trade-offs for vision tasks~\cite{hatamizadeh2025mambavision}.

Recent CD models have started adopting these backbones. ChangeMamba integrates VMamba into a bi-temporal architecture~\cite{chen2024changemamba}, and HAM-CD adds hybrid attention to improve fine-grained local modeling~\cite{li2026hamcd}. Wijenayake et al.\ explore spectral fusion and change-guided attention for semantic CD in Mamba-FCS~\cite{mambafcs}, and precise spatio-temporal fusion for binary CD~\cite{precision_spatiotemporal_fusion_iciis2025}. Mamba-CD~\cite{peng2026mambacd} is the closest related work, using change region-aware attention and gated recursive context refinement. In contrast, \ours{} constructs temporal evidence from paired features, absolute difference, and signed difference before decoding, then separates the fused representation into region and boundary streams through \drbi{}. CRAM-lite is used only as a lightweight residual modulation on D-RBI region features, while the main temporal decomposition is performed by \drbi{}.

These works show that SSM backbones are promising for remote-sensing CD, but most focus on region-level detection and do not explicitly decouple boundary-sensitive cues from region evidence.

\subsection{Boundary-Aware Modeling and Temporal Feature Construction}
Boundary precision is a long-standing challenge in dense prediction. Boundary loss penalizes contour errors directly~\cite{kervadec2019boundary}, Gated-SCNN uses a dedicated shape stream~\cite{takikawa2019gated}, and BASNet follows a coarse-to-fine boundary-aware design~\cite{qin2019basnet}. In remote sensing, boundary-aware multi-task learning has been studied for aerial dense prediction~\cite{wang2021boundary}, and BSCNet uses boundary-aware semantic context for sharper CD edges~\cite{zhou2026bscnet}.

Most CD methods compare bi-temporal features mainly through absolute differencing $|F_2-F_1|$. This captures change magnitude but discards temporal direction, so appearance and disappearance can become indistinguishable. Wasalathilaka et al.\ also report that boundary delineation remains a dominant failure mode for SSM backbones under domain shift~\cite{wasalathilaka2026controlled}. \ours{} addresses these issues by preserving signed temporal evidence through $F_2-F_1$ and separating the temporal descriptor into region and boundary pathways before decoding.
\section{Methodology}
\label{sec:method}

\subsection{Overview}
\ours{} predicts a binary change map from a bi-temporal pair $(I_1,I_2)$. As shown in Fig.~\ref{fig:architecture}, the model follows four steps. A shared MambaVision encoder extracts paired multi-scale features, D-RBI builds temporal evidence and separates it into region and boundary streams, ARF-FPN decodes the region stream into a coarse logit map, and a bounded residual head uses the boundary stream to refine uncertain contours. Thus, region features determine the main changed area, while Sobel-conditioned boundary features provide a controlled correction signal.

\subsection{Shared MambaVision Encoding}
\label{sec:encoder}
Both images pass through the same patch embedding and shared MambaVision encoder:
\begin{equation}
    \{F_1^\ell\}_{\ell=1}^{4}=E(I_1), \qquad
    \{F_2^\ell\}_{\ell=1}^{4}=E(I_2),
\end{equation}
where $\ell$ indexes the four encoder stages at scales $\{1/4,1/8,1/16,1/32\}$. In Fig.~\ref{fig:architecture}, each stage output from the first image is paired with the same-scale output from the second image. Weight sharing keeps both images in one feature space, so the following temporal differences mainly represent scene change rather than encoder mismatch. We use pretrained MambaVision-S as the default trainable backbone~\cite{hatamizadeh2025mambavision}.

\subsection{Differential Region--Boundary Interaction}
\label{sec:drbi}
At each scale $\ell$, D-RBI receives the paired features $F_1^\ell$ and $F_2^\ell$. The differential feature construction block in Fig.~\ref{fig:architecture} forms four streams:
\begin{equation}
    X^\ell =
    \left[
    F_1^\ell,\;
    F_2^\ell,\;
    |F_2^\ell-F_1^\ell|,\;
    F_2^\ell-F_1^\ell
    \right].
\end{equation}
The first two terms preserve the original temporal features, the absolute difference gives change magnitude, and the signed difference keeps temporal direction. The concat node $C$ stacks these streams along the channel dimension. The projection block then compresses the stacked evidence:
\begin{equation}
    D^\ell = \phi_\ell(X^\ell),
\end{equation}
where $\phi_\ell(\cdot)$ is the scale-specific $1\times1$ convolution shown in the diagram. The resulting differential feature $D^\ell$ is the shared descriptor from which both region and boundary responses are generated.

\textbf{Region gate.}
The upper branch in D-RBI computes a region gate from $D^\ell$:
\begin{equation}
    G_r^\ell = \sigma\!\left(\psi_\ell^r(D^\ell)\right), \qquad
    R_\ell = G_r^\ell \odot D^\ell .
\end{equation}
Here, $\psi_\ell^r(\cdot)$ denotes the lightweight convolutional gate and $\odot$ denotes element-wise multiplication. This equation means that the learned gate selects the parts of $D^\ell$ useful for changed-region localization. The output $R_\ell$ is the red region feature sent to the ARF-FPN decoder in Fig.~\ref{fig:architecture}.

\textbf{Boundary gate.}
The lower branch computes a boundary gate from a feature-level Sobel response:
\begin{equation}
    S^\ell = \operatorname{Sobel}(D^\ell), \qquad
    G_b^\ell = \sigma\!\left(\psi_\ell^b(S^\ell)\right).
\end{equation}
The Sobel operator is applied to the differential feature $D^\ell$, not to the raw image. Thus, the edge cue is tied to temporal change evidence rather than ordinary image texture. The boundary feature is
\begin{equation}
    B_\ell = G_b^\ell \odot D^\ell .
\end{equation}
This is the blue boundary feature in Fig.~\ref{fig:architecture}. Each D-RBI block therefore outputs a region feature $R_\ell$ for decoding and a boundary feature $B_\ell$ for residual refinement.

\subsection{CRAM-lite}
\label{sec:cramlite}
CRAM-lite is a lightweight residual spatial modulation block applied to the region stream at the first three encoder scales. Inspired by change-region-aware attention in Mamba-CD~\cite{peng2026mambacd}, it predicts a spatial map from the D-RBI region feature and applies residual gating:
\begin{equation}
    \widetilde{R}_{\ell}=R_{\ell}\cdot(1+\alpha A_{\ell}),
\end{equation}
where $A_{\ell}\in[0,1]$ is the learned modulation map and $\alpha$ is initialized to $0.5$. In the diagram, this block refines the region feature before decoding. CRAM-lite does not replace D-RBI; the main temporal decomposition is still performed by region-boundary separation.

\subsection{Adaptive Receptive-Field Decoder}
\label{sec:decoder}
The decoder operates on the CRAM-lite refined region stream. Since CRAM-lite is applied only to the first three scales, we set $\widetilde{R}_{4}=R_{4}$. The decoder input is $\{\widetilde{R}_{\ell}\}_{\ell=1}^{4}$. Each scale is processed by an adaptive receptive-field block with parallel dilated depthwise convolutions:
\begin{equation}
    d \in \{1,2,4,8\}.
\end{equation}
These branches give the decoder both local detail and wider spatial context. The multi-scale ARF-FPN block in Fig.~\ref{fig:architecture} fuses the four region features in a top-down manner, and the coarse prediction head produces
\begin{equation}
    P_c =
    h_{\mathrm{coarse}}\!\left(
    \operatorname{ARF\mbox{-}FPN}(\{\widetilde{R}_{\ell}\}_{\ell=1}^{4})
    \right),
\end{equation}
where $P_c$ is the coarse change logit map. This map is the region-level prediction before boundary correction.

\subsection{Bounded Boundary Residual Refinement}
\label{sec:refinement}
The refinement branch uses the selected finest-scale boundary feature $B_1$, the coarse logit $P_c$, and an edge cue from the coarse prediction:
\begin{equation}
    E_c = \operatorname{Sobel}(\sigma(P_c)).
\end{equation}
In Fig.~\ref{fig:architecture}, $E_c$ is the Sobel block applied to the coarse logits after sigmoid. The boundary aggregation block concatenates the boundary feature, coarse logit, and coarse edge cue:
\begin{equation}
    \Delta P = h_{\mathrm{ref}}\left([B_1,\;P_c,\;E_c]\right).
\end{equation}
The refinement head predicts a residual correction rather than a new mask. The final logit is
\begin{equation}
    P_f = P_c + 0.1\,\tanh(\Delta P).
\end{equation}
The $\tanh$ bounds the residual and the scale factor keeps the update small, so boundary evidence refines the coarse prediction without overwriting it. The final probability map is
\begin{equation}
    \hat{Y}=\sigma(P_f),
\end{equation}
which is thresholded to obtain the binary change map.

\subsection{Training Objective}
\label{sec:loss}
The full model is trained with final, coarse, and boundary supervision:
\begin{equation}
    \mathcal{L} =
    \mathcal{L}_{\mathrm{BCE}}
    + \mathcal{L}_{\mathrm{Dice}}
    + 0.4\,\mathcal{L}_{\mathrm{coarse}}
    + 0.1\,\mathcal{L}_{\mathrm{boundary}}.
\end{equation}
BCE and Dice supervise the final prediction $P_f$, while the coarse loss supervises $P_c$ to stabilize the region decoder. The boundary term applies a Sobel-edge $\ell_1$ loss to encourage contour consistency. The auxiliary weights keep the main BCE and Dice objective dominant, and ablations without full supervision disable the coarse and boundary losses.
\section{Experiments}
\label{sec:experiments}

\subsection{Datasets and Protocols}
\label{sec:datasets}
We evaluate \ours{} on DSIFN-CD~\cite{shi2022dsamnet} and WHU-CD~\cite{ji2019whu}. DSIFN-CD contains high-resolution bi-temporal image pairs from Chinese cities. Our verified DSIFN protocol uses explicit split files with 2\,758/394/789 train/validation/test images; the test set is evaluated as 3\,156 deterministic tiles. For WHU-CD, the verified project split contains 6096 training, 762 validation, and 762 test patches. Evaluation uses 256$\times$256 patches and the validation-selected threshold of 0.55.

\subsection{Implementation and Metrics}
Models are trained with AdamW, a cosine schedule, 2\,500 warmup iterations, batch size 8, 256$\times$256 crops, gradient clipping at 0.5, AMP, and EMA inference. Training uses random horizontal and vertical flips only. The full model uses MambaVision-S unless stated otherwise. For the verified DSIFN result, the 50k-iteration EMA checkpoint is evaluated at threshold 0.5. For WHU-CD, the verified threshold is 0.55. We report changed-class Precision (Pre), Recall (Rec), F1, IoU, and Overall Accuracy (OA). F1 and IoU are emphasized because OA is dominated by unchanged pixels.

\subsection{Boundary Metrics}
\label{sec:boundary_metrics}
BF1 is computed by extracting binary boundaries from the predicted and ground-truth masks and matching boundary pixels within a fixed tolerance. The boundary evaluation script uses a 3-pixel tolerance for BF1. BIoU is computed as the IoU between boundary bands obtained around the predicted and ground-truth contours with the same 3-pixel band. Trimap F1-3px computes changed-class F1 only inside a three-pixel band around the ground-truth boundary.

\subsection{Results and Literature Comparison}
\label{sec:sota_comparison}
Tables~\ref{tab:sota_dsifn} and~\ref{tab:sota_whu} compare \ours{} with compact sets of representative methods. Literature values are taken from the cited recent comparison table~\cite{peng2026mambacd}, parameter counts are shown only when verified. Colored numbers mark the best, second-best, and third-best values within each metric column.

\begin{figure*}[t]
    \centering
    \setlength{\tabcolsep}{1pt}

    \begin{tabular}{cc}
        \includegraphics[width=0.49\textwidth]{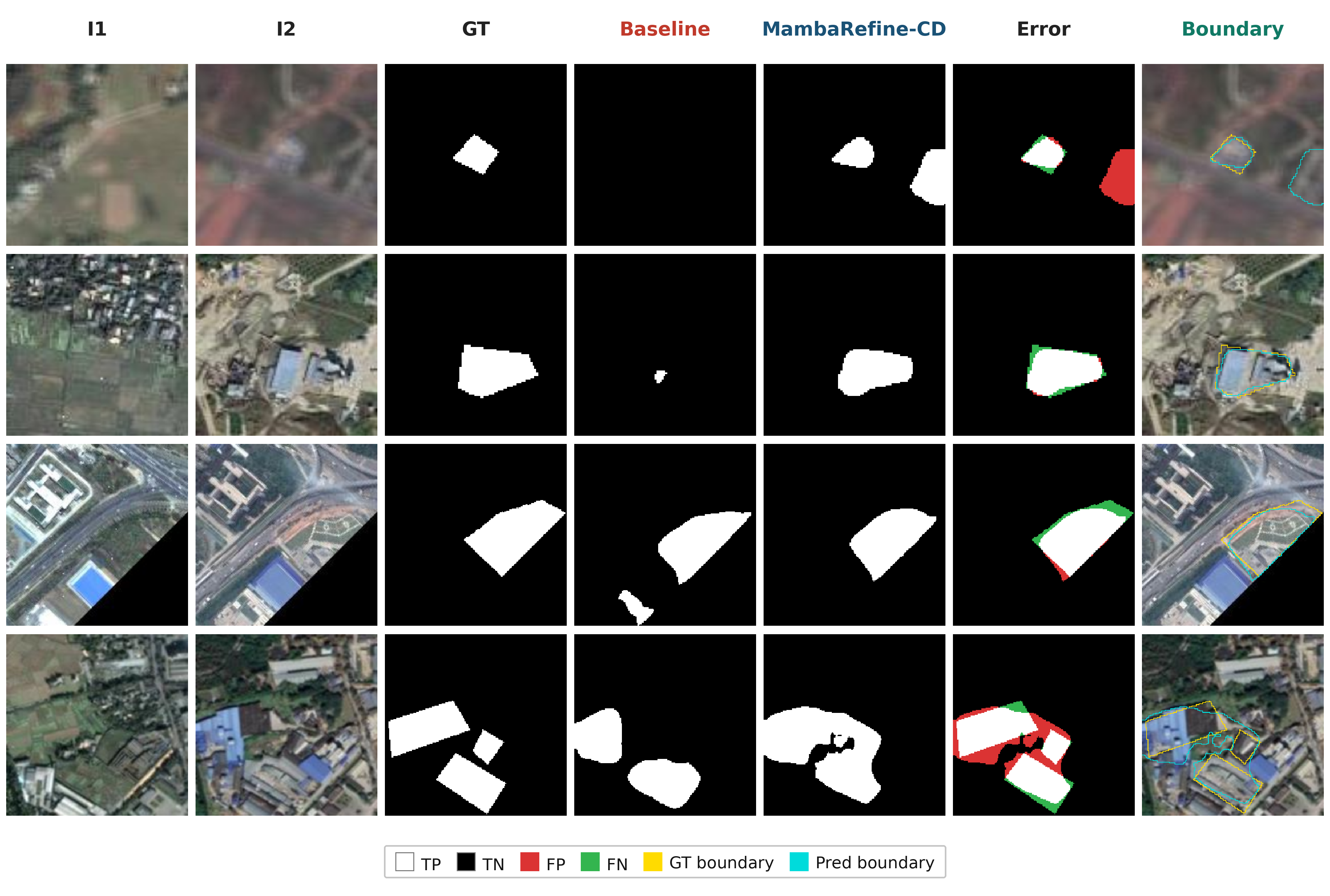}
        &
        \includegraphics[width=0.49\textwidth]{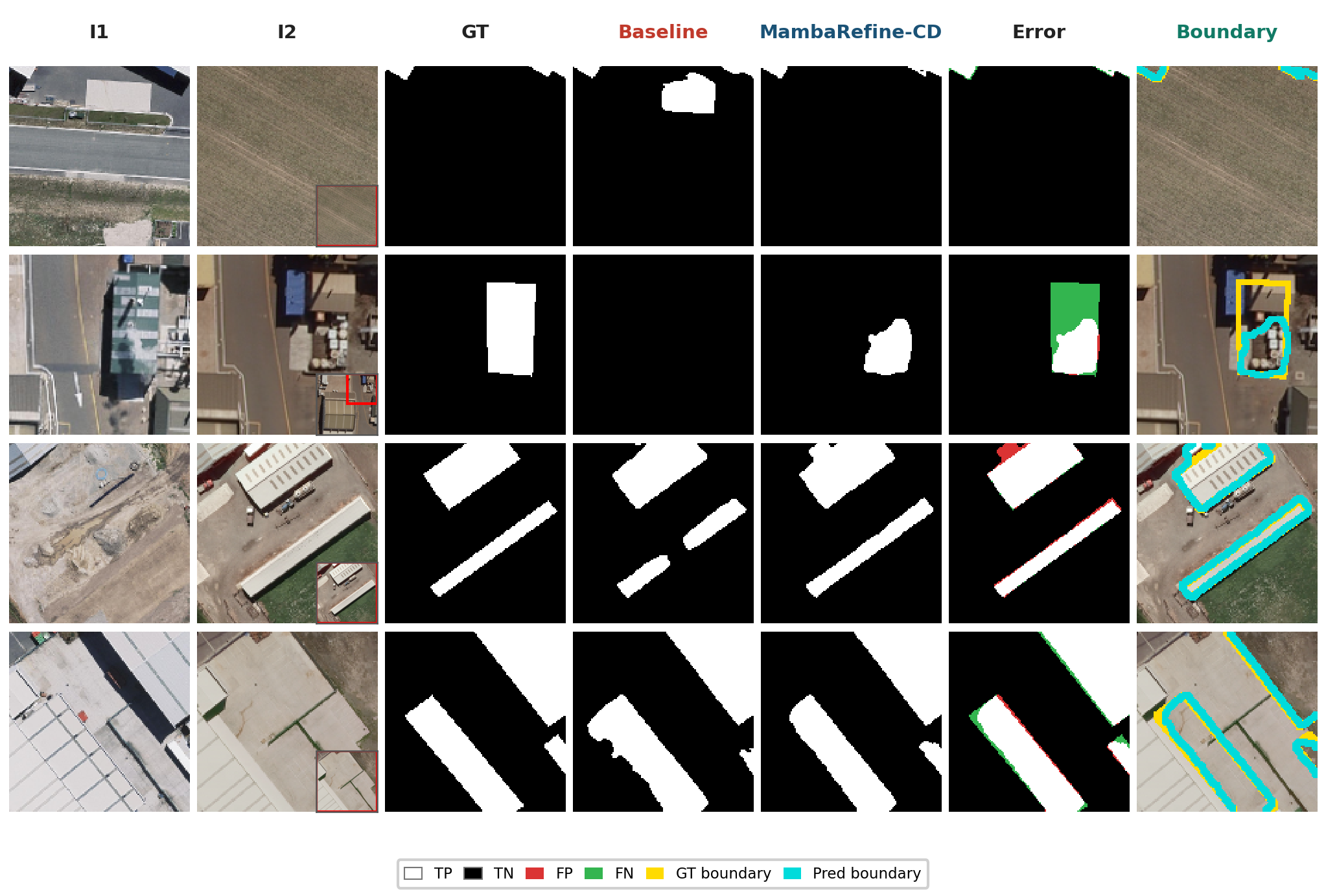}
        \\
        {\footnotesize (a) DSIFN-CD}
        &
        {\footnotesize (b) WHU-CD}
    \end{tabular}

    \caption{Qualitative results on DSIFN-CD and WHU-CD. Each panel shows the bi-temporal input pair, ground truth, prediction, and error map. White, black, red, and green denote true positives, true negatives, false positives, and false negatives, respectively.}
    \label{fig:qualitative}
\end{figure*}

On DSIFN-CD, \ours{} achieves 95.67\% F1 and 91.71\% IoU on average across three canonical full-model runs. These results are comparable to the Mamba-CD values reported by Peng et al.~\cite{peng2026mambacd}, which are 95.61\% F1 and 91.69\% IoU. However, because DSIFN-CD protocols differ across papers, this comparison should not be interpreted as a definitive same-protocol improvement. We therefore emphasize the verified split details and changed-class metrics, while reporting OA for completeness.

Peng et al.'s Mamba-CD~\cite{peng2026mambacd} is a closely related recent Mamba-based change detection method. On WHU-CD, \ours{} achieves 95.34\% F1 and 91.10\% IoU on average across three runs, compared with 95.20\% F1 and 90.83\% IoU reported by Mamba-CD. This indicates competitive changed-region overlap, while Mamba-CD retains higher precision and OA. The best single WHU-CD run of \ours{} further reaches 95.53\% F1 and 91.45\% IoU, suggesting that the proposed region-boundary refinement design is competitive with recent Mamba-based methods.

\subsection{Ablation Study}
\label{sec:ablation}
Table~\ref{tab:ablation} reports clean-split DSIFN ablations. The MambaVision encoder gives the largest gain over the SimpleCNN baseline. Unsigned D-RBI alone gives only a small and unstable change, while signed temporal difference improves the D-RBI design. ARF-FPN gives a modest positive gain. The boundary residual head alone does not improve the model by itself. The full model improves after combining boundary residual refinement with CRAM-Lite, coarse auxiliary supervision, and boundary loss this supports the full region-boundary refinement pipeline, not every individual module independently. Single-run ablations are reported where repeated runs were unavailable.

\subsection{Boundary Evaluation}
\label{sec:boundary_evaluation}

Table~\ref{tab:boundary_analysis} reports the boundary metrics defined in Section~\ref{sec:boundary_metrics}. Its region metrics come from the checkpoints used for boundary evaluation, while Table~\ref{tab:ablation} reports the main ablation runs, so A1 and A4 may differ slightly across the two tables.

The full MambaRefine-CD pipeline improves BF1 from 61.36\% to 71.94\% and BIoU from 38.41\% to 47.39\% over the MambaVision-FPN baseline. This supports the proposed region-boundary refinement design. The A5 result further shows that boundary residual refinement is most effective when combined with signed temporal modeling, ARF-FPN decoding, CRAM-lite modulation, and boundary-aware supervision.

\begin{table}[t]
\centering
\caption{Contextual DSIFN-CD comparison. Literature values are from Peng et al.~\cite{peng2026mambacd}; rankings are not strict same-split SOTA claims because DSIFN-CD split protocols differ across papers.}
\label{tab:sota_dsifn}
\scriptsize
\setlength{\tabcolsep}{2.0pt}
\begin{tabular*}{\columnwidth}{@{\extracolsep{\fill}}lccccc@{}}
\toprule
Model & Pre & Rec & F1 & IoU & OA \\
\midrule
FC-Siam-Conc~\cite{daudt2018fully}          & 66.45 & 54.21 & 59.71 & 42.56 & 87.57 \\
SNUNet~\cite{fang2022snunet}                & 60.60 & 72.89 & 66.18 & 49.45 & 87.34 \\
ChangeFormer~\cite{bandara2022changeformer} & 88.48 & 84.94 & 86.67 & 76.48 & 95.56 \\
BiFA                                        & 73.99 & 68.87 & 71.34 & 55.45 & 90.80 \\
FTAN                                        & 90.54 & 88.61 & 89.56 & 81.10 & -- \\
ADSFNet                                     & \second{94.79} & \second{95.24} & \second{95.01} & \second{90.50} & \second{98.30} \\
Mamba-CD~\cite{peng2026mambacd}             & \best{95.60} & \third{95.61} & \third{95.61} & \third{91.69} & \best{98.51} \\
\midrule
\ours{}                                     & \third{95.47} & \best{95.87} & \best{95.67} & \best{91.71} & 96.98 \\
\bottomrule
\end{tabular*}
\end{table}

\begin{table}[t]
\centering
\caption{Ablation on DSIFN-CD clean split. A0--A6 use 50K iterations.}
\label{tab:ablation}
\scriptsize
\setlength{\tabcolsep}{1.4pt}
\resizebox{\columnwidth}{!}{%
\begin{tabular}{c l l c c c c}
\toprule
ID & Variant & Main Added Component & Params (M) & F1 & IoU & OA \\
\midrule
A0 & FPN Baseline        & SimpleCNN encoder        & 7.84  & 76.63 & 62.12 & 83.80 \\
A1 & MambaVision-S + FPN & MambaVision encoder      & 53.54 & 93.21 & 87.28 & 95.72 \\
A2& + D-RBI unsigned & Absolute-diff D-RBI & 54.98 & 93.33 & 87.50 & 95.35 \\
A3 & + Signed Diff       & Signed temporal stream   & 55.34 & 94.28 & 89.19 & 95.99 \\
A4 & + ARF-FPN           & Adaptive RF decoder      & 65.12 & 94.36 & 89.32 & 96.05 \\
A5 & + Boundary Residual & Bounded residual head    & 65.19 & 93.59 & 87.94 & 95.53 \\
A6 & Full Model          & CRAM-Lite + aux/bnd.\ loss & 65.40 & \best{95.67} & \best{91.71} & \best{96.98} \\
\bottomrule
\end{tabular}
}
\end{table}

\begin{table}[t]
\centering
\caption{WHU-CD comparison with representative methods. Literature values are from Peng et al.~\cite{peng2026mambacd}; our result is the average of three full-model EMA runs.}
\label{tab:sota_whu}
\scriptsize
\setlength{\tabcolsep}{2.0pt}
\begin{tabular*}{\columnwidth}{@{\extracolsep{\fill}}lccccc@{}}
\toprule
Method & Pre & Rec & F1 & IoU & OA \\
\midrule
FC-EF~\cite{daudt2018fully}                  & 71.63 & 67.25 & 69.37 & 53.11 & 97.61 \\
STANet                 & 79.37 & 85.50 & 82.32 & 69.95 & 98.52 \\
SNUNet~\cite{fang2022snunet}                 & 85.60 & 81.49 & 83.50 & 71.67 & 98.71 \\
IFNet                  & \best{96.91} & 73.19 & 83.40 & 71.52 & 98.83 \\
ChangeFormer~\cite{bandara2022changeformer}  & 91.83 & 88.02 & 89.88 & 81.63 & 99.12 \\
BiFA                             & 95.15 & \third{93.60} & \third{94.37} & \third{89.34} & \second{99.56} \\
RSM-CD                 & 93.37 & 90.42 & 91.87 & 84.96 & -- \\
SChanger                    & 94.62 & 91.83 & 93.20 & 87.27 & -- \\
CDMamba                     & 95.58 & 92.01 & 93.76 & 88.26 & 99.51 \\
Mamba-CD~\cite{peng2026mambacd}              & \second{96.52} & \second{93.91} & \second{95.20} & \second{90.83} & \best{99.62} \\
\midrule
\ours{}                                      & \third{95.79} & \best{94.90} & \best{95.34} & \best{91.10} & \second{99.56} \\
\bottomrule
\end{tabular*}
\end{table}

\begin{table}[t]
\centering
\caption{Boundary-aware evaluation of MambaRefine-CD variants using BF1, BIoU and 3-pixel Trimap F1.}
\label{tab:boundary_analysis}
\resizebox{\columnwidth}{!}{
\begin{tabular}{lccccc}
\hline
\textbf{Variant} & \textbf{F1} & \textbf{IoU} & \textbf{BF1} & \textbf{BIoU} & \textbf{Trimap F1$_{3px}$} \\
\hline
A1 MambaVision-FPN        & 93.21 & 87.28 & 61.36 & 38.41 & 68.02 \\
A4 + ARF-FPN              & 93.71 & 88.16 & 65.94 & 42.06 & 70.27 \\
A5 + Boundary Residual    & 93.59 & 87.94 & 65.14 & 41.37 & 69.98 \\
A6 Full Model             & \textbf{95.67} & \textbf{91.71} & \textbf{71.94} & \textbf{47.39} & \textbf{72.93} \\
WHU Full Model            & 95.15 & 90.76 & 89.49 & 71.37 & 87.01 \\
\hline
\end{tabular}
}
\end{table}
\subsection{Backbone Scaling}
\label{sec:backbone_scaling}

Table~\ref{tab:backbone_variants} compares backbone scaling. MambaVision-B provides only marginal gains over MambaVision-S with substantially more parameters, so MambaVision-S is used as the main backbone, while tiny and base variants show the accuracy-efficiency trade-off.

\begin{table}[t]
\centering
\caption{Parameters and GFLOPs are reported for the full MambaRefine-CD model using each MambaVision backbone variant.}
\label{tab:backbone_variants}
\scriptsize
\setlength{\tabcolsep}{3.0pt}
\begin{tabular}{lccccc}
\toprule
Backbone & Params (M) & GFLOPs & F1 & IoU & OA \\
\midrule
MambaVision-T & 46.80  & 50.05  & 94.58 & 89.72 & 96.24 \\
MambaVision-S & 65.40  & 64.08  & 95.67 & 91.71 & 96.98 \\
MambaVision-B & 113.44 & 109.15 & 95.77 & 91.89 & 97.07 \\
\bottomrule
\end{tabular}
\end{table}

\subsection{Qualitative Results}
\label{sec:qualitative}

Fig.~\ref{fig:qualitative} provides a visual comparison on DSIFN-CD and WHU-CD. The examples highlight changed-region completeness, boundary consistency, false positives, and false negatives.

\subsection{Discussion and Limitations}
\label{sec:discussion}

The DSIFN-CD and WHU-CD results show that \ours{} achieves strong changed-class F1 and IoU under verified evaluation settings while remaining competitive with recent Mamba-based methods. The main contribution is not only the final score, but the region-boundary temporal refinement design. \drbi{} separates region evidence from Sobel-conditioned boundary evidence, and the bounded residual branch refines uncertain boundary logits without dominating the main prediction.

The DSIFN-CD literature comparison should be interpreted conservatively because published works may use different patch-level splits. Table~\ref{tab:sota_dsifn} is therefore a contextual comparison rather than a strict same-split SOTA ranking.

The boundary evaluation in Table~\ref{tab:boundary_analysis} supports the role of the full region-boundary refinement pipeline. However, direct boundary-metric comparison against Mamba-CD is not reported because official prediction masks or checkpoints were not available in our experimental setup. Future work will evaluate learned boundary priors, perform same-protocol baseline re-evaluation, and extend the analysis to additional CD benchmarks.
\section{Conclusion}
\label{sec:conclusion}

This paper presented \ours{}, a MambaVision-based binary change detection model with explicit region-boundary refinement. The method combines signed temporal evidence in \drbi{}, ARF-FPN decoding, CRAM-lite modulation, and bounded residual correction for boundary refinement.

Experiments on DSIFN-CD and WHU-CD show strong changed-class F1 and IoU under verified protocols, with competitive performance against recent Mamba-based methods. Ablations indicate that the full combination of signed temporal modeling, auxiliary supervision, boundary loss, and residual refinement is more effective than using the boundary branch alone.

\bibliographystyle{IEEEtran}
\bibliography{references}

\end{document}